\documentclass{iopart}

\usepackage{subfigure,cite,hyperref}
\usepackage{color}
\usepackage{graphicx}

\newcommand{\be}{\begin{equation}}
\newcommand{\ee}{\end{equation}}
\newcommand{\beq}{\begin{eqnarray}}
\newcommand{\eeq}{\end{eqnarray}}

\newcommand{\f}{\frac}

\begin{document}

\title[]{Semiclassical states in quantum gravity: Curvature associated to a Voronoi graph}

\author{Jacobo D\'iaz-Polo$^1$ and I\~naki Garay$^2$}

\address{$^1$ Department of Physics and Astronomy, Louisiana State
University, Baton Rouge, LA, 70803-4001.}

\address{$^2$ Programa de P\'os-Gradua\c{c}\~ao em F\'isica, Universidade Federal do Par\'a, 66075-110, Bel\'em, PA, Brazil.}

\eads{\mailto{jacobo@phys.lsu.edu}, \mailto{inakigaray@ufpa.br}}

\date{\today}

\begin{abstract}
The building blocks of a quantum theory of general relativity are expected to be discrete structures. Loop quantum gravity is formulated using a basis of spin networks (wave functions over oriented graphs with coloured edges), thus realizing  the aforementioned expectation. Semiclassical states should, however, reproduce the classical smooth geometry in the appropriate limits. The question of how to recover a continuous geometry from these discrete structures is, therefore, relevant in this context. Following previous works by Bombelli \emph{et al.} we explore this problem  from a  rather general mathematical perspective using, in particular, properties of Voronoi graphs to search for their compatible continuous geometries.  We test the previously proposed methods for computing the curvature associated to such graphs and analyse  the framework in detail, in the light of the results obtained.
\end{abstract}






\section{Introduction}

General relativity describes the gravitational interaction as a consequence of the curvature of space-time (a 4-dimensional Lorentzian manifold). Given this geometric nature of gravity, it is expected that, when quantizing, a prescription for quantum geometry would arise based on more fundamental discrete structures, rather than on smooth differential manifolds \cite{Isham:1995wr}.

Loop quantum gravity (LQG) is a candidate theory for such a quantization of general relativity \cite{thiemann2007modern}. The fundamental objects (basis of the kinematical Hilbert space) are the so-called spin network states, which are defined as wave functions constructed over oriented coloured graphs. The building blocks of the theory are, therefore, discrete combinatorial structures (graphs). The theory provides quantum operators with a direct geometric interpretation (areas and volumes), which happen to have discrete eigenvalues, reinforcing the idea of a discrete geometry \cite{Rovelli:1994ge,Ashtekar:1996eg,Ashtekar:1997fb,Dittrich:2007th}.

This perspective of considering abstract combinatorial structures as the fundamental objects of the theory is also adopted in other approaches to quantum gravity, such as spin-foam models \cite{Perez:2012wv}, causal dynamical triangulations \cite{Loll:1998aj} and causal sets \cite{Henson:2006kf}. Also, the algebraic quantum gravity approach \cite{Giesel:2006uj} follows the same spirit of constructing a quantum theory of gravity from an abstract combinatorial structure.

Despite the variety of successful results obtained in LQG, the search for a semiclassical sector of the theory that would connect with the classical description given by general relativity in terms of a smooth manifold is still under research. An interesting question for the description of a semiclassical sector, given the combinatorial nature of the building blocks of the theory, would be whether there is any correspondence between certain types of graphs and the continuous classical geometries. Tentatively, this would allow for the construction of gravitational coherent states corresponding to solutions of Einstein field equations. While it is certainly true that spin networks are a particular basis, and coherent states constructed from them might resemble nothing like a graph, some works \cite{Freidel:2010aq,Freidel:2010bw,Livine:2011vk,Dupuis:2010iq} seem to indicate that these graphs do actually represent the structure of space-time at the fundamental level.

This raises a very interesting question. How does the transition between a fundamental discrete geometry, encoded in a graph structure, and the continuous geometry we experience in every-day life happen? In particular, how does a one-dimensional structure give rise to 3-dimensional smooth space? A step towards answering these questions could be to think of these graphs as \emph{embedded} in the corresponding continuous geometries they represent. However, the situation is rather the opposite, being the smooth continuous structure an effective structure, emerging from the more fundamental discrete one, and not the other way around. Therefore, a very relevant question to ask would be: Is there any information, contained in the abstract structure of a graph, that determines (or restricts) the compatible continuous geometries? Can we determine what types of manifolds (dimensionality, topology) a certain graph can be embedded in? One could even go further and ask whether any additional geometric information (like curvature) can be extracted from the very abstract structure of the graph itself. The goal is, therefore, to construct a unique correspondence between the discrete structures given by graphs, which in general do not carry geometric information, and smooth manifolds. This problem was studied in the context of quantum gravity by Bombelli, Corichi and Winkler \cite{Bombelli:2004si,Bombelli:2009bb}. Almost a decade ago, they proposed a statistical method to compute the curvature of the manifold that would be associated to a certain class of graphs, based on Voronoi diagrams, giving a new step towards the semiclassical limit of LQG. Indeed, due to their properties, Voronoi diagrams appear naturally when addressing this kind of problems and have been used in this context before \cite{Thiemann:2002vj,Ashtekar:2001xp,Bombelli:2000ua,Oriti:2012us,Calcagni:2012cv}. They also play an important role in the discrete approach to general relativity provided by Regge calculus \cite{McDonald:2008js,Miller:1997wb,McDonald:2008gm}.

Voronoi diagrams are generated from a metric manifold and, by construction, contain geometric information from it. What was proposed in \cite{Bombelli:2004si}, however, is to \emph{throw away} all additional geometric information and to keep only the abstract structure of the one-dimensional graph that forms the skeleton of the Voronoi diagram. Then, the task is to study if there are any imprints of the original geometry which remain in this abstract graph structure. Although the work is somewhat preliminary and, for the most part, restricted to 2-dimensional surfaces\footnote{There are also similar studies made for surfaces in 3 dimensions with negative curvature \cite{isokawa}.}, it tackles very interesting questions and explores a novel path towards a semiclassical regime in LQG. The results obtained could also provide a useful tool for the causal dynamical triangulations approach \cite{Henson:2009fy}.

Our goal in this paper is to explicitly implement this mechanism to compute the curvature of Voronoi graphs generated from 2-dimensional surfaces, with the aim of testing the validity of the approach in \cite{Bombelli:2004si} and checking how accurate the obtained results are. To that end, we carried out a computer-based implementation, generating random Voronoi diagrams with standard algorithms, then \emph{stripping} those diagrams from all geometric information, and finally applying the proposed method to obtain a curvature that we subsequently compare to the original surface used to construct the diagram. While doing so, we analyse and discuss the difficulties encountered in producing estimations of the curvature which, ultimately, seem to render the studied method ineffective. We explore several alternatives to overcome those difficulties and suggest some directions for further research.

The structure of the paper is the following. In section \ref{section1} we review the mathematical setting and definitions relevant to this framework and describe in more detail the method proposed in \cite{Bombelli:2004si} for the computation of the curvature of Voronoi graphs. Section \ref{section2} describes the computational implementation of this method and the analysis of the obtained results, while introducing additional considerations about the topology of the sampled subgraphs. In section \ref{section3}, a closer look at some statistical details related to the sampling procedure is presented, including issues that are at the heart of the difficulties found to effectively use this approach. Finally, in section \ref{section4}, we summarize the results and present the conclusions of this work.

\section{Curvature associated to a graph}\label{section1}

In this section we give a brief introduction to the mathematical setting surrounding the problem that we want to study with special attention to the description and motivation for the use of Voronoi graphs, which play a central role in this framework. After this, we describe the method proposed by Bombelli, Corichi and Winkler \cite{Bombelli:2004si} to compute the curvature associated to such graphs.

\subsection{Mathematical setting and Voronoi graphs}

\begin{figure}[ht]
\begin{center}
\includegraphics[width=4.5cm]{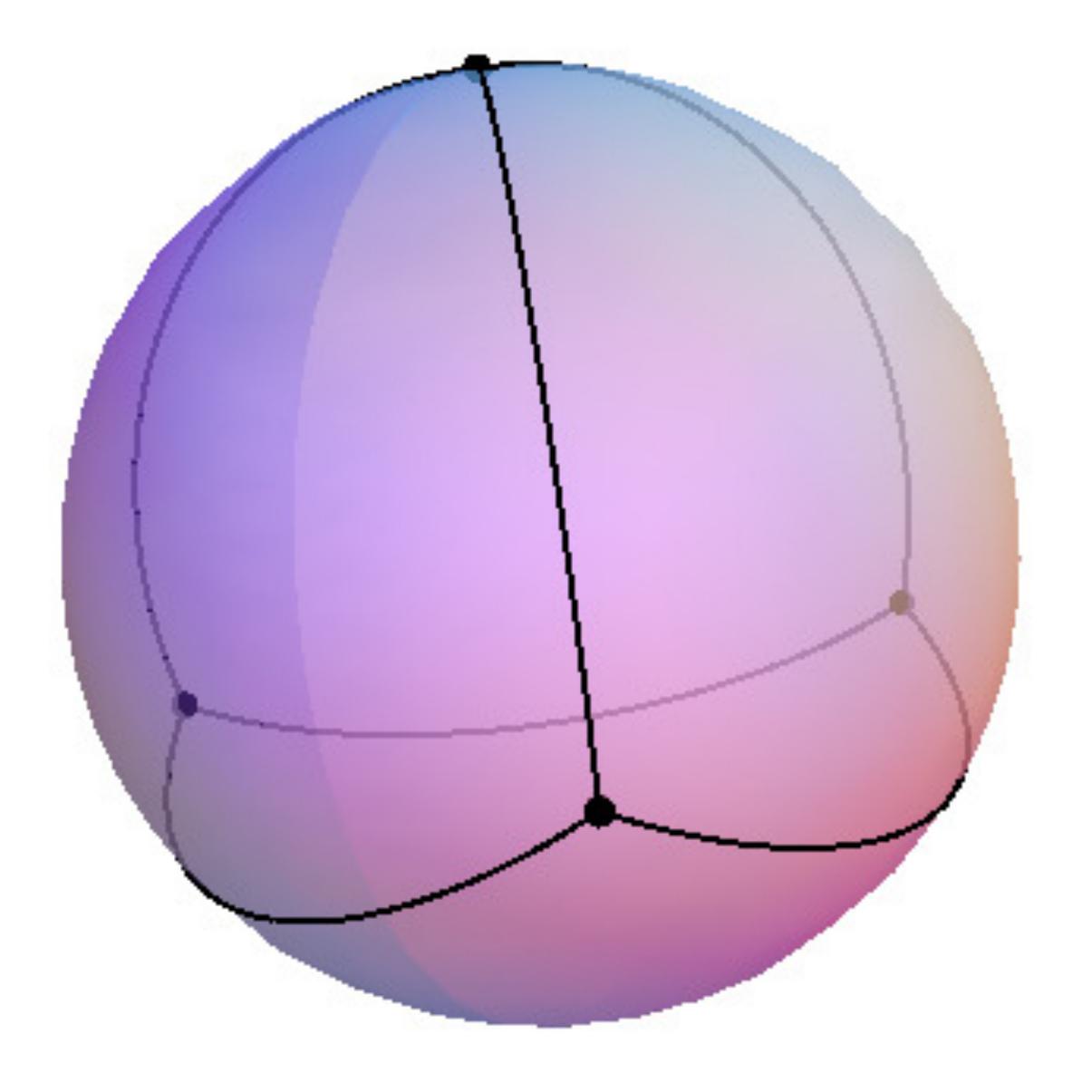}
\hspace*{1cm}
\includegraphics[width=4.5cm]{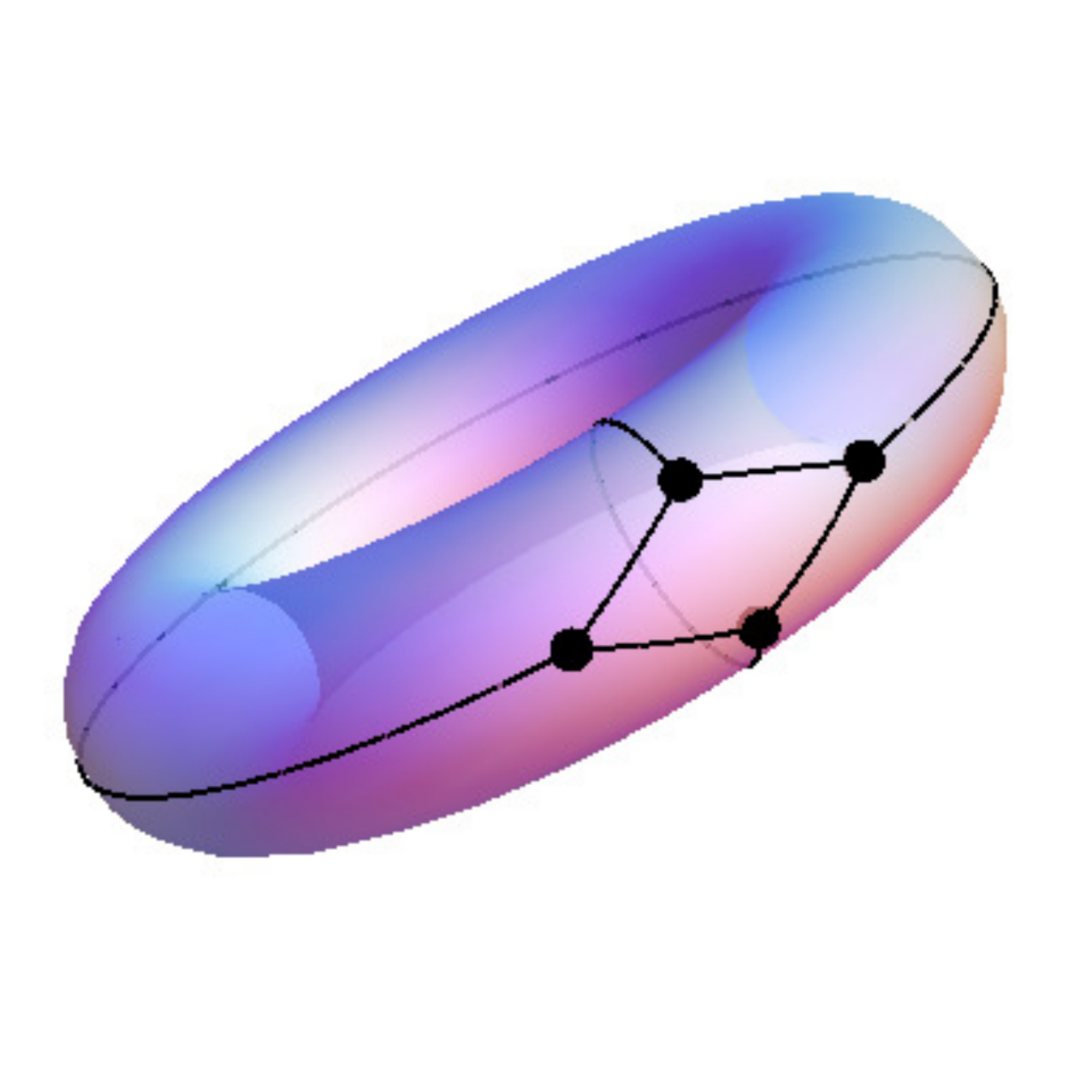}
\caption{These two figures represent a tetrahedron embedded both in a sphere and in a torus. Note that the key is choosing a set of closed loops in the abstract graph representing the corresponding faces on the surface. Two different choices (sets of four and two faces) are shown in this case, resulting in the spherical and toroidal topologies. This is a trivial example of the non-uniqueness of the solution to the problem of embedding a given graph into a manifold with arbitrary topology. See \cite{Kramer:2002xn}  for more details on the embedding of graphs into surfaces.
\label{tetraedro}}
\end{center}
\end{figure}

The mathematical question of associating a manifold to a given abstract graph is a hard one already at the topological level (it is an undecidable problem). For instance, given a tetrahedral graph, we could associate it to (embed it in) either a sphere or a torus (see figure \ref{tetraedro}). On the other hand, this problem constitutes an example of the so-called inverse problems \cite{Smolin:2005mq}, and one of the features of this kind of problems is the non-uniqueness of their solution \cite{tarantola2005inverse}. As a consequence, we need to provide additional information (constraint the initial data and/or \emph{a priori} select certain class of results) in order to have a unique solution. 

The inverse problem studied here arises in several approaches to quantum gravity that display a discrete structure at the fundamental level. In the case of LQG this correspondence would appear, in particular, between a graph and a 3-dimensional spatial manifold. Assumedly, this graph would be useful to construct a semiclassical state for the theory. There is a specific type of graphs, Voronoi graphs \cite{okabe2009spatial,aurenhammer1996voronoi}, which appear naturally in this context. In the spirit of \cite{Bombelli:2004si}, we restrict the study presented here exclusively to this kind of graphs. This is the first piece of additional information (restriction) we introduce in order to solve the problem (more restrictions are discussed in the next section).

A Voronoi diagram is constructed in the following way. For a set of points (seeds) on a metric space, each highest-dimensional cell of the Voronoi diagram contains only one seed, and comprises the region of space closer to that one seed than to any of the others. Then, co-dimension $n$ cells are made by sets of points equidistant to $n+1$ seeds, e.g., in 2 dimensions, the edges (1-dimensional cells) of the Voronoi diagram are the lines separating two of these regions, and are therefore equidistant to two seeds. In the same way, vertices (0-dimensional cells) are equidistant to three seeds. Therefore, except in degenerate situations---which are avoided by randomly sprinkling the seeds---, the valence of all vertices in a $D$-dimensional Voronoi diagram is $D+1$. Another interesting property of Voronoi diagrams is that their dual graph is the so-called Delaunay triangulation, whose vertices are the Voronoi seeds. By construction, for a given set of seeds on a metric space the corresponding Voronoi diagram is uniquely defined.

\begin{figure}[ht]
\begin{center}
\includegraphics[width=5.5cm]{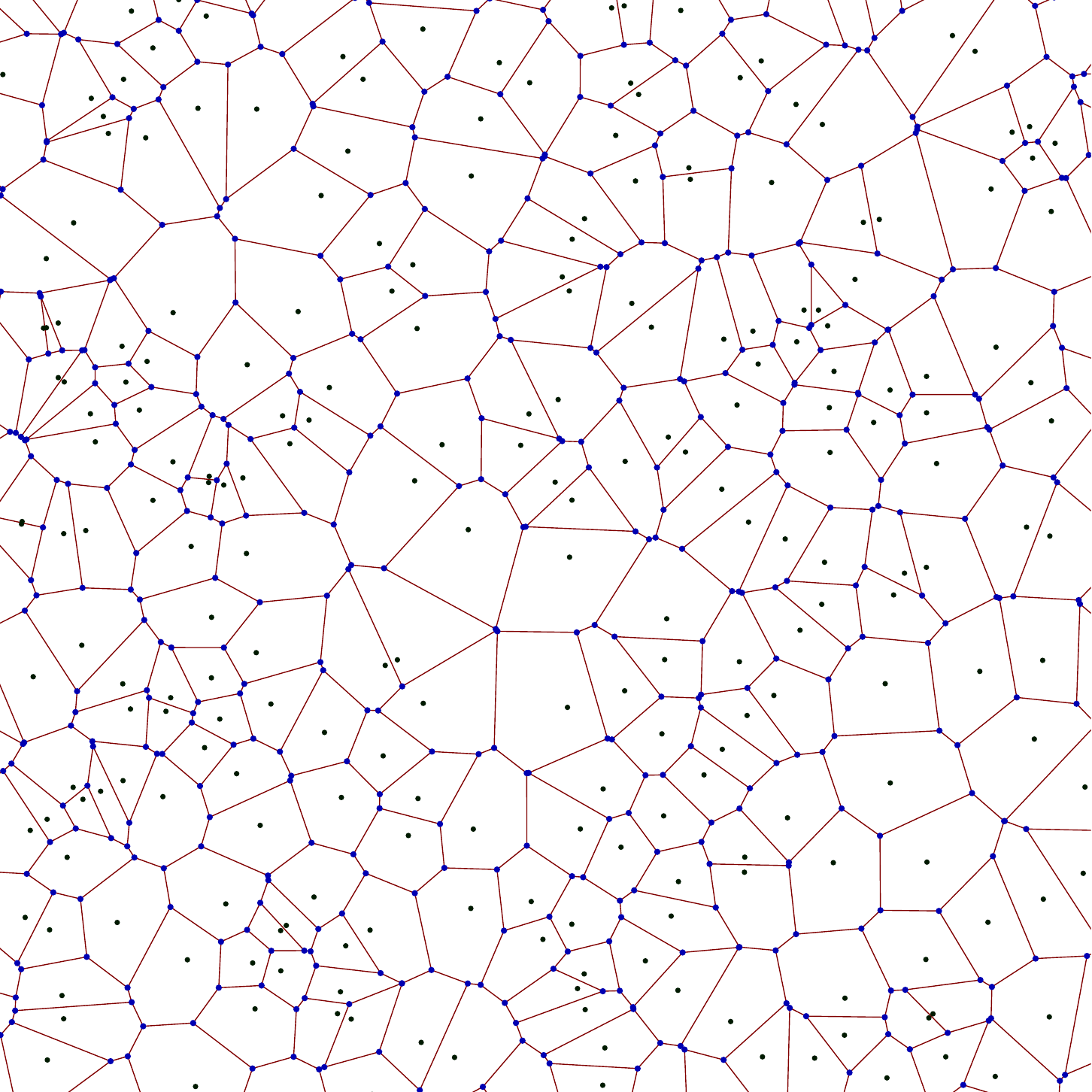}
\hspace*{0.5cm}
\includegraphics[width=5.5cm]{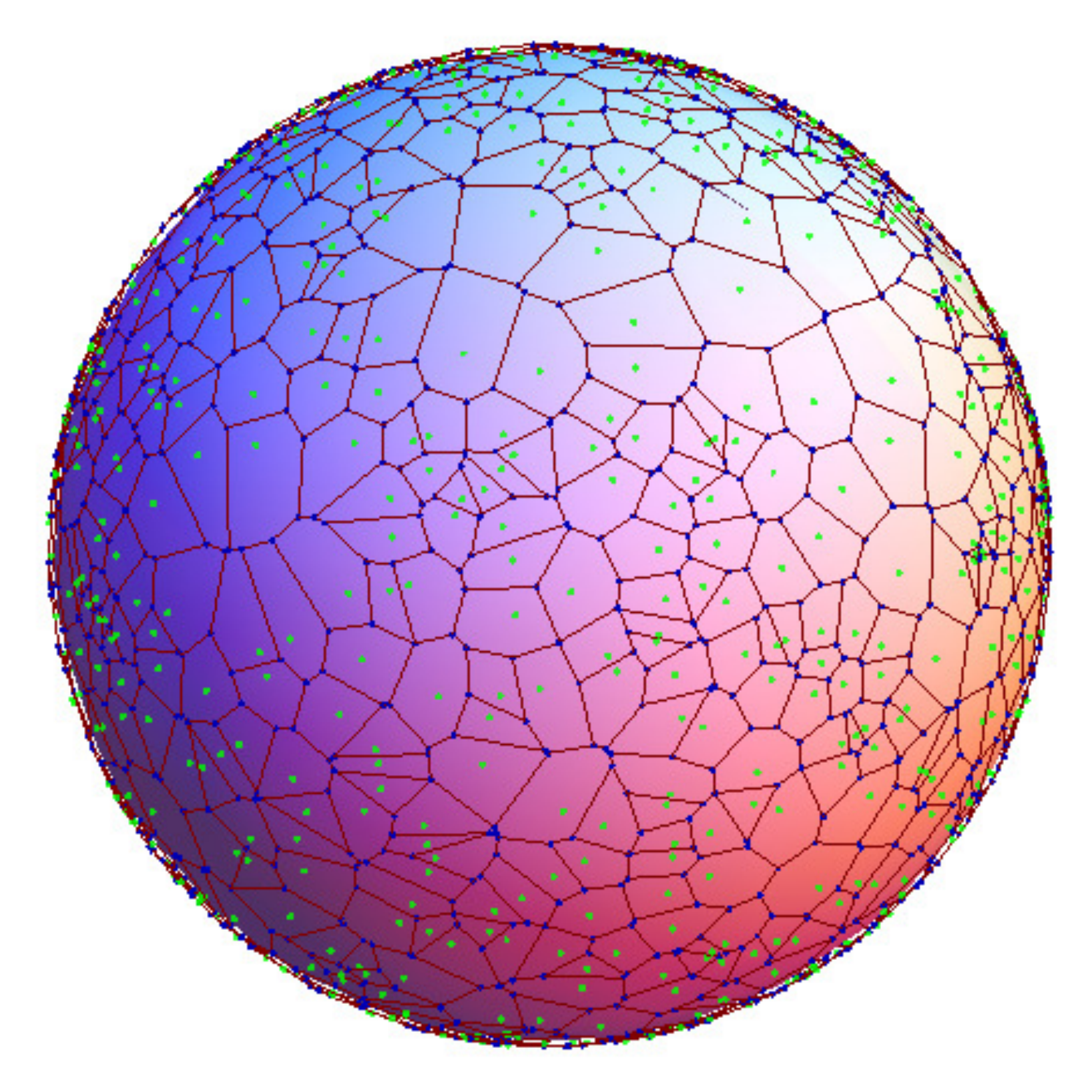}
\caption{Planar and spherical Voronoi diagrams computed for a random distribution of seeds.
\label{voronois}}
\end{center}
\end{figure}

In this paper we use standard algorithms \cite{Joe1991325,Renka:1997} to construct Voronoi diagrams corresponding to random distributions of seeds both on the unit sphere and in a plane (see figure \ref{voronois}). The graphs generated this way will be the test bed for the computations performed henceforth.

\subsection{Description of the method}

We review, in this section, the main part of \cite{Bombelli:2004si,Bombelli:2009bb} that we will be working with. The reader can refer to the original papers for a more complete exposition. Here we focus, in particular, on their 2-dimensional proposal. The starting point is to consider a given surface on which we randomly sprinkle a set of points, that will be the seeds for the Voronoi construction. A Voronoi cell-complex\footnote{A set of non empty pairwise disjoint cells such that (a) The closure of each cell is homeomorphic to a ball and its boundary homeomorphic to a sphere, and (b) The boundary of each cell is a union of cells \cite{Bombelli:2004si}.} 
is constructed, containing zero, one, and two-dimensional cells (vertices, edges and faces). We keep, then, the abstract structure of the one-dimensional graph (encoded, for instance, in an adjacency matrix) and forget the rest of the information. We are, thus, left with an abstract graph.

Does that abstract graph structure ``remember'' anything about the surface it was created from? The first step to test this is to try to recover the two-dimensional structure of the diagram, i.e., assign faces to the abstract graph. In \cite{Bombelli:2004si} this question is tackled by defining a \emph{plaquette} as a closed loop in the graph such that the shortest path in the graph between any two vertices contained in the loop is part of the loop. The idea is then to recover the original cell-complex starting from the abstract graph by applying this definition. However, as also pointed out there, that definition by itself is not enough to recover the original structure. One needs to impose additional conditions, namely, that every edge is shared by two and only two \emph{plaquettes}. This requirement, however, derives from \emph{a priori} imposing that the obtained diagram be 2-dimensional. This once again is a consequence---as pointed out above---of the fact that we are dealing with an inverse problem with non-unique solutions.

\begin{figure}[ht]
\begin{center}
\includegraphics[width=5cm]{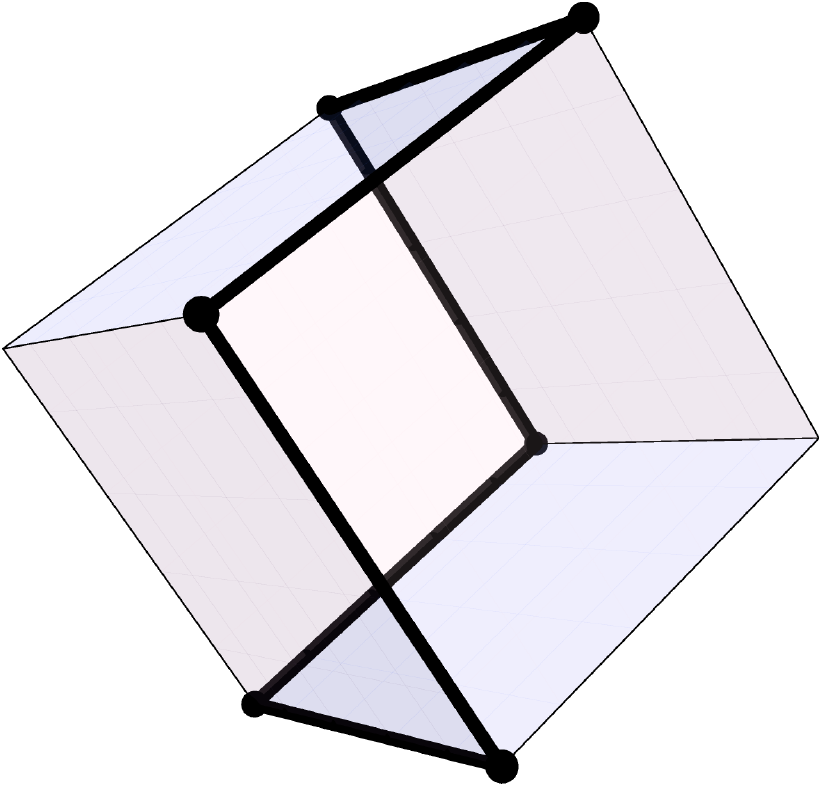}
\caption{The highlighted loop satisfies the definition of \emph{plaquette} in this cubic graph, but it is not one of the faces of the cube. Restricting every edge to be shared by only the two smallest \emph{plaquettes} containing it, excludes this ``false face'' from being considered.
\label{cube}}
\end{center}
\end{figure}

We have been able to explicitly confirm these facts for particular examples. By applying only the definition of \emph{plaquette}, many more, bigger loops than the original faces are obtained. A very simple example of that fact can be seen, for instance, in figure \ref{cube}. The highlighted loop satisfies the definition of \emph{plaquette}, however it is clearly not one of the faces of the 2-dimensional cell-complex formed by the cube. When one restricts the search for only the smallest (in number of sides) two \emph{plaquettes} per edge, the original Voronoi cell-complex is recovered. Having checked this fact, we will skip the step of undoing and redoing the cell-complex in the computations of curvature for large Voronoi diagrams, for the sake of computational simplicity. In the remainder of this paper, we analyse 2-dimensional Voronoi cell-complexes, including the information on which sets of vertices form the different faces. Note, however, that we will not keep any distances or geometrical quantities associated to the cell-complex, but only the abstract structure---the information that we would be able to recover, anyway, from just the abstract Voronoi graph.

Once the two-dimensional diagram has been reconstructed, the task is to compute the curvature. In order to do so, one can use certain relations between the number of faces, edges and vertices in the diagram. Since the sprinkling of points that gave rise to the Voronoi diagram was chosen at random, we can assume that degenerate situations in which a point of the manifold is equidistant to four or more seeds do not occur. In that case, all vertices of the Voronoi graph are tri-valent. This gives rise to the following relation between the total number $V$ of vertices in the graph and the total number of edges $E$:
\be
\label{tri-valent}
V=\frac{2}{3}E,
\ee
since every vertex is shared by three edges and every edge contains two vertices.
We will also use the definition of the Euler-Poincar\' e characteristic $\chi$ in the two-dimensional case
\be
\label{Euler-Poincare}
\chi=V-E+F,
\ee
where $F$ is the total number of faces in the graph (that equals the number of seeds).
Finally, one can define the number $p$ of sides of a face (its perimeter in the graph). Taking into account that every edge is shared by two faces, the average $\bar p$ over a set of faces satisfies
\be
\label{average}
\bar p=\frac{2E}{F}.
\ee
From (\ref{tri-valent}), (\ref{Euler-Poincare}), and (\ref{average}) the following expression for the Euler-Poincar\' e characteristic $\chi$ can be obtained
\be
\label{chi}
\chi=F\left(1-\frac{\bar p}{6}\right),
\ee
in terms of the total number of faces $F$ and the average number of sides of the faces $\bar p$. 

On the other hand, if there is a manifold $M$ associated to the Voronoi diagram, this manifold should have the same topology (thus the same $\chi$) as the diagram. The Gauss-Bonnet theorem can be used then to relate $\chi$ with the integral of the curvature over the manifold. If $M$ is a manifold without boundary (like a sphere), the theorem takes the form
\be
\label{Gauss-Bonnet}
\chi=\frac{1}{4\pi}\int_{M} R\ dA,
\ee
where $dA$ is the area measure and $R$ is the Ricci scalar.

At this point Bombelli \emph{et al.} introduce an important assumption. Equation (\ref{Gauss-Bonnet}) only holds for the integral over the whole manifold, since $\chi$ is a topological quantity. However, one can assume that the region of the graph one is looking at is small enough so that the curvature can be considered constant. In that case---and assuming positive curvature---one would be looking at a patch of a sphere. Now, it is possible to make the abstraction that there actually is a full sphere containing that region of the graph, and for that sphere the Gauss-Bonnet theorem applies. But, if that were true, it would mean that one can integrate out the constant curvature of that sphere and obtain the following expression:
\be
\label{BCW1}
R=4\pi\frac{F}{A_S}\left(1-\frac{\bar p}{6}\right),
\ee
where $A_S$ is the total surface area of the sphere. However, if we assume a uniform distribution of faces (a uniform distribution of seeds to generate the Voronoi diagram), then this formula can also be applied to the sphere patch by defining a density of faces  $\rho=F/A_S=F_P/A_P$, where $F_P$ and $A_P$ are respectively the number of faces and area of the patch. If the graph is fine enough so that the sphere patch that we are considering contains a large enough number of faces to constitute a good statistical sample, then the average of $p$ computed over the patch should be the same as the average for the whole sphere. Following this reasoning, the formula to compute the curvature of a small graph region of constant curvature, in terms of $\rho$ and $\bar p$, is proposed in \cite{Bombelli:2004si}:
\be
\label{BCW2}
R=4\pi\rho\left(1-\frac{\bar p}{6}\right).
\ee
In order to obtain a value of curvature with this formula, one needs to introduce a scale in the problem by providing a value of $\rho$. This is a direct consequence of the fact that the abstract Voronoi graph is invariant under re-scalings of the sphere radius. Notice, however, that fixing $\rho$ here does not fix the area $A_S$ of the whole sphere---which would trivially determine the curvature---but only the area $A_P$ of the patch. In a sense, the only non-trivial information one can aspire to obtain, given the `scale invariance' of the abstract Voronoi graph, is the fraction of an imaginary sphere that would be covered by the considered patch. We aim to do that by statistically determining the average $\bar p$.

The question, now, is how to use the previous formula in an effective way to apply to a graph and obtain a sensible value of curvature. The idea is to find the best compromise between a large enough region, so that statistics are good, and a small enough region, so that the constant curvature approximation holds. In order to determine what the ideal size for that region is, Bombelli \emph{et al.} consider the sequence of \emph{layers} $\lambda_n$ of $n$-degree neighbours of a certain initial face. Every time a new layer of neighbours is added to the previously-selected faces, one can compute the dispersion (standard deviation) of $p$ over the region $\sum_{i=0}^n \lambda_i$. As the region size increases, it is expectable that this dispersion will decrease, as a consequence of having a better statistical sample. However, at some point, the region will be too large for the curvature to be constant, and the change of curvature within the region will cause the dispersion to start increasing again. Whenever the minimum dispersion is reached, one can assume that the best compromise has been achieved. For that size of the region, the average $\bar p$ is computed, and the curvature obtained from it is considered the best estimate for the patch. Of course, this procedure is adaptable to different situations, and the size of the ideal region is naturally adapted to how fast the curvature changes, allowing to get much larger regions whenever the curvature is slow-varying.

The procedure summarized above is the main result we want to test in this paper. In the next section, our goal will be to explicitly implement that procedure and compute the curvature of surfaces for particular cases.


\section{Implementation and results}\label{section2}


Here we carry out a computer-based implementation of the procedure described in the previous section for the simplest geometries in 2 dimensions: the unit sphere and the plane. We analyse the results and their accuracy for different levels of graph refinement. We then comment on possible boundary effects arising from the disk topology of the sampled regions and study the consistency of the method in the light of those considerations.

\subsection{Testing the method}

In order to test the method described above, the first thing we need is an efficient way to generate Voronoi diagrams. Fortunately, Voronoi diagrams are very useful tools for a variety of fields (ecology, image processing, military purposes, networks...). Because of that, very efficient algorithms have already been developed and are available. We make use of standard algorithms to generate Voronoi diagrams in a sphere \cite{Renka:1997} and in a plane \cite{Joe1991325}.

Our first step will be to focus on the simplest possible case: a sphere. A sphere is convenient both from the computational perspective---since we know its curvature at every point and standard algorithms to generate the Voronoi diagrams are available---as well as from the perspective of the practical implementation, as one does not have to worry about the considered region being too large. For a sphere, the constant-curvature approximation always holds; therefore, the larger the region, the better. This makes a sphere the perfect scenario for our first check, aimed at testing whether this method can be used to compute a sensible approximate value of its curvature. We should be able to observe the value of curvature tending to the right one, and stabilizing around it as the region increases in size, at the same time as the dispersion decreases.

\begin{figure}[ht]
\begin{center}
\includegraphics[width=0.49\textwidth]{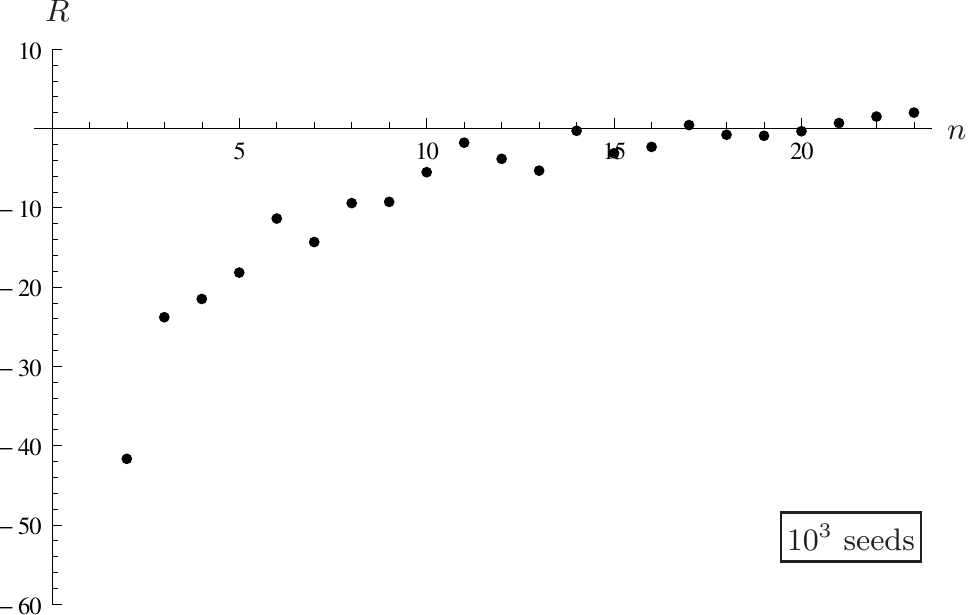}
\hfill
\includegraphics[width=0.49\textwidth]{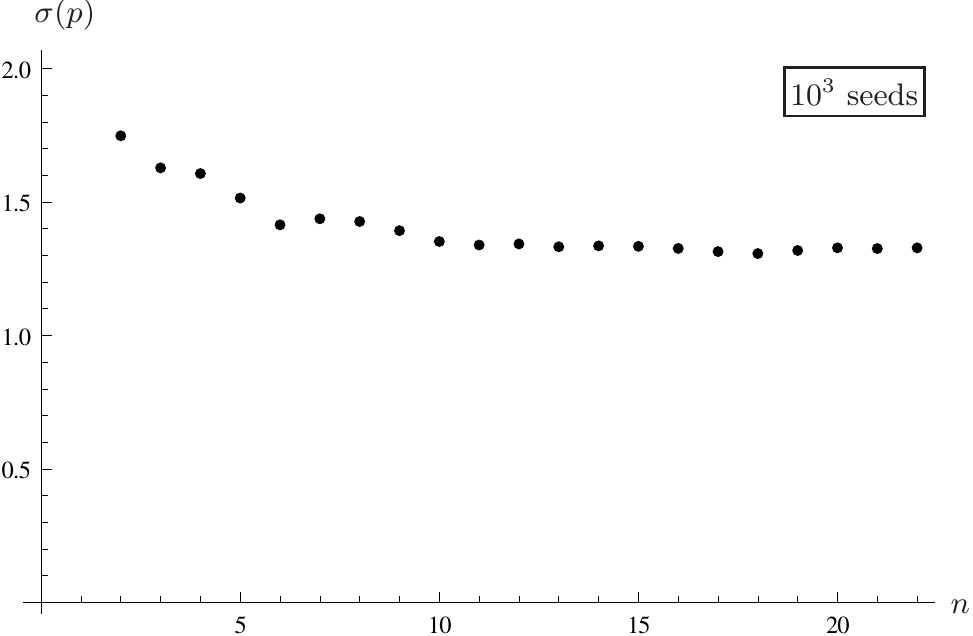}
\caption{The computed values for the Ricci scalar, on the left, and the standard deviation $\sigma(p)$, on the right, are represented as a function of the number $n$ of layers $\lambda_i$ considered. Computations are performed for a Voronoi diagram generated from a collection of $10^3$ seeds randomly distributed over the surface of a sphere.}
\label{1Kcell}
\end{center}
\end{figure}

With that goal in mind, we generate Voronoi diagrams corresponding to random sprinklings of points on the unit-radius sphere (taking care of the fact that the random sprinkling is done in a curved geometry \cite{Bombelli:2000ua}). We start with coarse graphs, generated with fewer seeds, and keep refining them by generating new Voronoi diagrams with more and more seeds. In each case, we randomly choose an initial face and start building the consecutive ``layered'' regions. At each step, following the prescription, we compute the corresponding curvature and the standard deviation $\sigma(p)$ in the considered set of faces. The results for the case of a Voronoi diagram generated on a sphere with $10^3$ seeds are shown in figure \ref{1Kcell}.

We can see several things going wrong there. On the one hand, the standard deviation does not decrease with the size of the region, but seems to stabilize relatively soon, around a value of $\sim\!\!1.3$. On the other hand, the curvature does not stabilize to any given value, but it keeps changing all the way until it reaches the right value ($R=2$ for the unit sphere) just at the very end, when the considered region is the whole sphere. One could ask, in principle, how important that variation is in relative terms. It might be just a small variation, representing the fact that the statistics keep getting better. However, there seems to be no room for that interpretation, for two reasons. First, it is not a variation that is randomly spread around the right value, and whose amplitude tends to decrease. Much on the contrary, curvature is a monotonically increasing function of the size of the region. And second, and most importantly: the curvature of the sphere is (for the most part) negative! So, there's no room for this to be an approximate enough value of the curvature. It is completely off, except in the case where the region encloses the whole sphere.

\begin{figure}[ht]
\begin{center}
\includegraphics[width=0.49\textwidth]{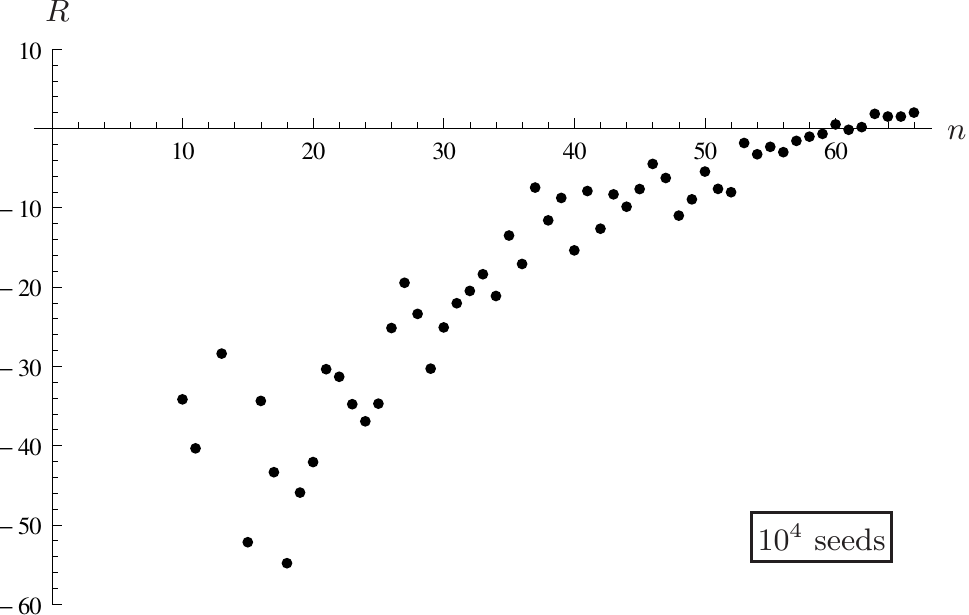}
\hfill
\includegraphics[width=0.49\textwidth]{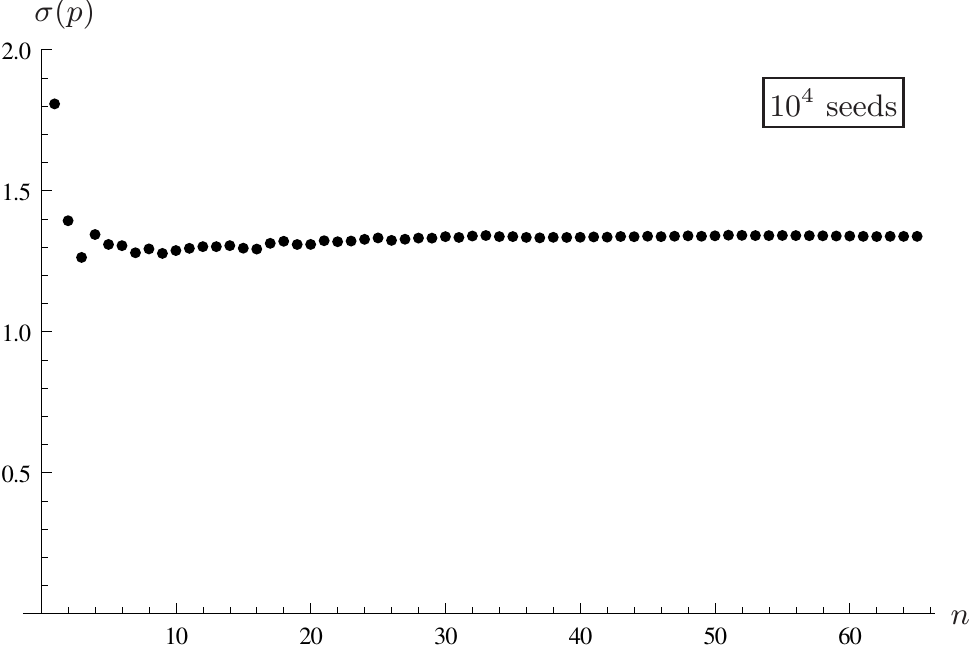}
\caption{As in figure \ref{1Kcell}, the curvature and the standard deviation $\sigma(p)$ are represented, this time for a Voronoi diagram generated using $10^4$ random seeds.
\label{10Kmediasdesv}}
\end{center}
\end{figure}

This, however, is a very coarse graph, and it was expectable that with only a thousand faces the statistics would not be good enough. We go ahead and repeat the calculations, this time taking $5\times 10^3$ seeds, $10^4$ seeds, $2\times 10^4$ seeds... The case with $10^4$ seeds is shown in figure \ref{10Kmediasdesv}.
What we observe while doing this is a little bit more worrying. We keep observing the exact same problem. The fact that we are refining the graph does not seem to be helping. Rather than having the curvature go to positive values for a region about the same size (number of layers) than in the $10^3$ case, and stabilizing around a good value thereafter, what we see instead is that, the finer the graph, the more negative the curvature gets, ``pushing away'' to the right the point where curvatures cross to positive values. Again, only for the very few last stages of the process, when the region covers almost the whole sphere, the values of curvature are positive. There seems to be a sort of ``scale invariance'' that appears to indicate that the only relevant parameter is the fraction of the total sphere that the region is covering, regardless of how fine the graph is (we will see in section \ref{statistics} that this is indeed the case). This behaviour makes sense, given the topological nature of some of the considerations put into the derivation of the used formula. However, that does not explain why the computed value for the curvature is completely off.

In order to further investigate this issue, we go to an even simpler case. We generate a Voronoi diagram on a flat, two-dimensional surface. With that we intend to simplify the problem to the maximum, getting rid of any possible unnoticed factors that might be causing the variation of curvature with the sample size. For a flat diagram, the curvature should always be zero (except fluctuations) making any undesired proportionality factors that might have sneaked in irrelevant. However, when we go ahead and do the exact same computations, we obtain similar results. Again, the curvature is off, not taking both positive and negative values spread around an average of zero, but taking only negative values all the way. This time, however, there is not a region size for which the whole manifold is contained in the sample and the topological arguments are exact. In the case of a flat surface, the sampled region is only limited by the finite size of the graph we are able to construct computationally. As a consequence, the curvature seems to get closer to zero as the region size increases, but it does not seem to ever reach the right value. On the contrary, it stabilizes, appearing to asymptotically tend to a constant negative value.

All this seems to indicate that there is something else affecting the computations and driving the curvature down to negative values. We consider, at this point, two possible aspects that can be the cause for this effect. The first one, in next section, are the  possible boundary effects due to the fact that the considered samples have the topology of a disk, and therefore have a boundary. The second aspect we will explore, in section \ref{friends}, is the possibility of some statistical bias introduced by the way in which the particular layered sampled regions are selected.

\subsection{Topology of the samples and boundary terms}

\begin{figure}[t]
\begin{center}
\includegraphics[width=9cm]{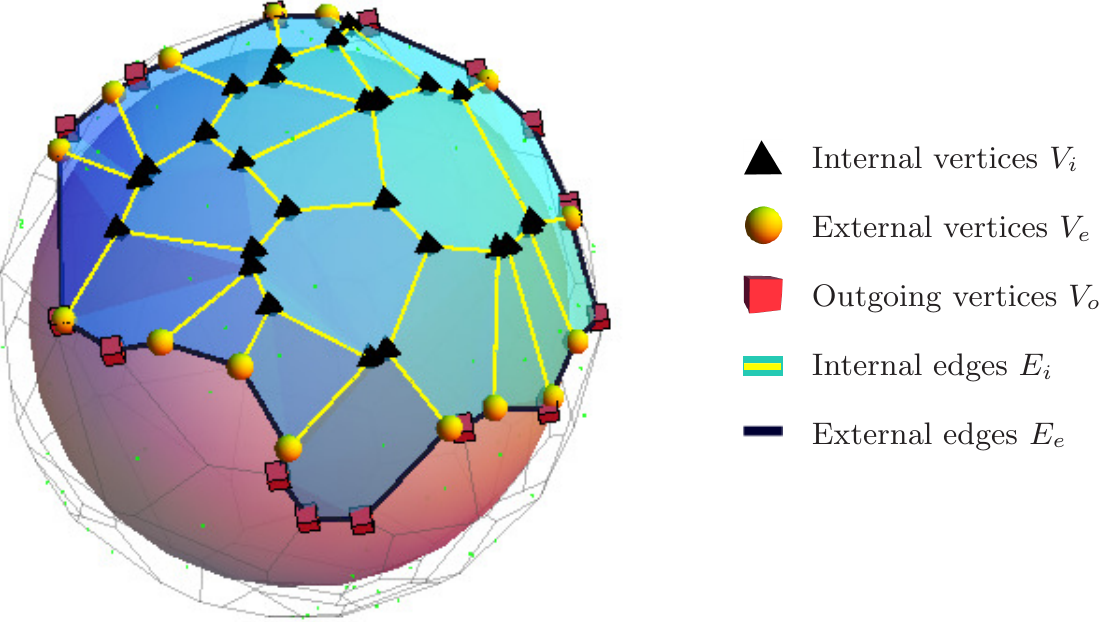}
\caption{The different types of edges and vertices in a sample region are represented in this figure.
\label{intext}}
\end{center}
\end{figure}

In order to study possible boundary effects, we take a more detailed look at the shape of the region we are considering for our computations. Although the argument in \cite{Bombelli:2004si} considers a manifold without boundary, therefore using only the bulk term in the Gauss-Bonnet theorem, in practice, the regions we are sampling over have disk topologies (sphere patches). This has several consequences, not only on the Gauss-Bonnet theorem---which acquires a boundary term---but on the relationships between total numbers of vertices, edges and faces that we are using. In particular, we have two different kinds of edges---internal and external ones. Internal edges share two faces, but external ones (the ones on the boundary) only form part of one face within the sampled region. Therefore, equation (\ref{average}) for the average of $p$ is modified, as the contribution of the total number of internal edges $E_i$ and the total number of external edges $E_e$ to the average is different:
\be
\label{average_disc}
\bar p=\frac{2E_i+E_e}{F_i},
\ee
where now $F_i$ is the number of faces inside the selected region.
Furthermore, we also have three different types of vertices, that we call \emph{internal}, \emph{external} and \emph{outgoing} (see figure \ref{intext}). Internal ones are inside the region, in the bulk. External and outgoing vertices are on the boundary. Internal vertices share three internal edges. External vertices, despite being on the boundary, still share three edges in the region. Two of those are external edges (on the boundary) and the third one goes inside the region, and is thus an internal one. Finally, outgoing vertices only share two edges in the region, both of them on the boundary. The third edge going out of outgoing vertices is outside the region (hence the nomenclature), and it is thus not considered. All this obviously modifies the relationship (\ref{tri-valent}) between number of vertices and number of edges, which now actually gets split in two equations: One for the total number $E_i$ of internal edges in terms of the number $V_i$ of internal vertices and the number $V_e$ of external indices, and another one for the total number $E_e$ of external edges in terms of the number $V_e$ of external vertices and the number $V_o$ of outgoing vertices:
\be
\label{internal external edges}
E_i=\frac{3}{2}V_i+\frac{1}{2}V_e\ , \quad \quad \quad E_e=V_e+V_o\ .
\ee
Substituting these new relations in (\ref{average_disc}), and using again (\ref{Euler-Poincare}), it is possible to arrive to a new equation, analogous to (\ref{chi}), but taking into account the disk topology of the samples:
\be
\label{chi_boundary}
\chi=F_i\left(1-\frac{\bar p}{6}\right) + \frac{V_o-V_e}{6}.
\ee

We observe that we get the same expression (\ref{chi}) we were getting before but with an additional term, which depends only on the boundary vertices. On the other hand, as mentioned above, now the Gauss-Bonnet theorem acquires also a boundary term:
\be
\label{Gauss-Bonnet_boundary}
\chi=\frac{1}{4\pi}\int_{M} R\ dA+\frac{1}{2\pi}\int_{\partial M}k_g ds,
\ee
where $\partial M$ is the boundary of the manifold $M$, and $k_g$ is the geodesic curvature. Now, equating expressions (\ref{chi_boundary}) and (\ref{Gauss-Bonnet_boundary}), we get, for a disk topology,
\be
\label{disc}
\frac{1}{4\pi}\int_{M} R\ dA+\frac{1}{2\pi}\int_{\partial M}k_g ds=F_i\left(1-\frac{\bar p}{6}\right) + \frac{V_o-V_e}{6}.
\ee
This formula is exact, and no assumptions need to be made for it to be satisfied. If we compute the right-hand side for a particular graph region, with a disk topology, the result is always $1$ (for a disk, $\chi=1$). In order to go forward and obtain a value of the curvature, we still need to make some assumptions. In particular, besides the constant-curvature approximation, now we also need to postulate a splitting between bulk and boundary terms on the right-hand side of (\ref{disc}), since only then we can isolate the value of the bulk integral on the left-hand side, which contains the curvature. As long as both terms, bulk and boundary, are added  together, the result is just a constant number (the Euler-Poincar\' e characteristic $\chi$).

Looking at the right-hand side of (\ref{disc}), though, there seems to be a natural splitting. The first term corresponds, as we saw, to the bulk term in the case of a sphere, and it depends on quantities on the bulk (the number of faces in the region $F_i$ and the average $\bar p$ of the number of sides of those faces), while the second term only appears when we consider a boundary and, as commented above, it only depends on boundary quantities (the number of boundary vertices, both outgoing $V_o$ and external $V_e$). It seems quite obvious and natural to make the straightforward splitting and associate the first term to the bulk and the second term to the boundary. However, if we do so, we just recover equation (\ref{BCW2}) for the curvature\footnote{Note that the manifold $M$ one integrates over on the l.h.s. of (\ref{disc}) is now the patch of the sphere corresponding to the selected region. Therefore, 
$
\frac{1}{4\pi}\int_{M} R\ dA=F_i\left(1-\frac{\bar p}{6}\right)\,\Rightarrow\,R=4\pi\f{F_i}{A_R}\left(1-\frac{\bar p}{6}\right)\,,
$
where $A_R$ is now the area of that region. Recalling that we are working with a uniform distribution of faces with density $\rho=F_i/A_R=F/A_S$, equation (\ref{BCW2}) is obtained.}, as given in \cite{Bombelli:2004si}.

Two conclusions can be drawn from this. First, if we are to find something different stemming from these new equations, such that allows us to correct the value of the obtained curvature, that would imply considering a splitting between bulk and boundary different from what appears to be the most natural one. Such a splitting should be justified, or at least motivated, on the basis of reasonable additional considerations. Second, as long as one sticks to the natural splitting, these results seem to confirm equation (\ref{BCW2}), and the additional boundary considerations made here seem to add some robustness to the arguments followed in its derivation.

Boundary effects do not seem to be, therefore, at the origin of the difficulties in finding a sensible curvature estimate. In the next section we look at the details of how the sampling regions are chosen.


\section{Statistical issues}\label{section3}

In the previous section we tested the method proposed by Bombelli \emph{et al.} for computing the curvature associated to a Voronoi graph both on a sphere and in a plane, and we obtained puzzling results. Our goal in this section is to understand more deeply the causes of the observed deviations and point out some relevant issues that should be taken into account when tackling this kind of problems.

\subsection{Statistical sampling and the \emph{friendship paradox}}\label{friends}

As we saw above, a very particular procedure has been suggested in order to carry out the computations and effectively use equation (\ref{BCW2}) to estimate the curvature of a graph. One starts with a single initial face, and then adds to it layers $\lambda_n$ of all $n$-degree neighbours. At each step $n$, computations are performed on the corresponding subgraph, made out of the faces contained in $\sum_{i=0}^n\lambda_i$. The question we want to ask here is, is there anything wrong with this procedure? In particular, is there any bias in the way of choosing new faces to expand the considered region, such that the computations could be affected?

\begin{figure}[ht]
\begin{center}
\includegraphics[width=5.5cm]{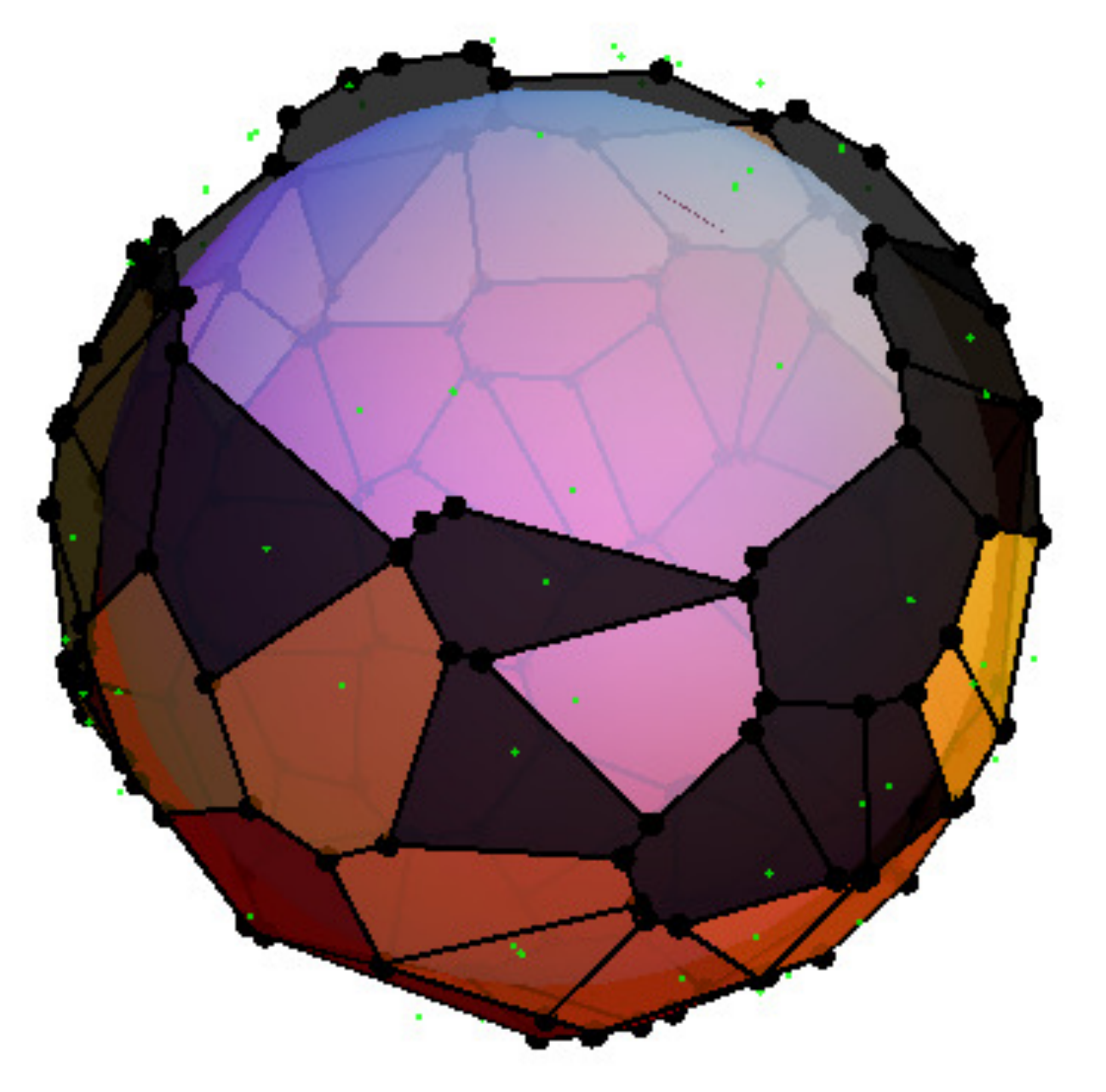}
\caption{In this plot we observe that a hole (an unselected face surrounded by selected ones) appears in the last layer of the sampled region (the darkest coloured one). Notice that, although there are some vertices that seem 4-valent, that is only due to the big size of the dots used to represent them for visual purposes. When 4 edges appear to be connected to a single vertex, there are indeed two nearby 3-valent vertices blended in the drawing.
\label{hole}}
\end{center}
\end{figure}

The first thing we can check is whether there is actually something funny going on with the topology of the region itself. One can, in principle, think of certain faces being larger than others, then advancing the progress of the region's expansion faster in some directions than in others with smaller faces. We cannot, therefore, expect the region to be circular in a precise sense. However, a deformed shape should not affect the topology-based considerations. There is another related issue worth looking at. If the expansion of the region is irregular in different directions, one could also think of situations in which two extremes of the boundary would close onto themselves, leaving a ``hole'' in the disk, made out of yet-unreached faces (see figure \ref{hole}). This situation would cause the topology of the sample to change, possibly  affecting the computation of the curvature. If holes are happening often, when the regions get large there could be an increasing number of them, requiring their careful consideration.

However, equation (\ref{chi_boundary}) is useful precisely to compute the topology of the region. We can just use that equation to see how often (and how much) $\chi$ deviates from $1$, and whether or not there is a correlation between those deviations and the deviations of curvature from the correct value. We performed those computations and found that, although the mentioned holes do happen occasionally, they are exceptions, and in most cases the topology of the region remains that of a disk. Furthermore, it is rare that more than one hole happen at the same time. And, in any case, there seems to be no correlation at all between holes happening and deviation of the curvature.

There is a more subtle issue, related to the choice of region, that could be affecting our computations. When using the concept of neighbouring faces as the basic structure for the sampling, one is bringing in the factor of the different connectivities (number of neighbours) of the faces. Different connectivities imply modifications on the probability distributions of faces, and might very well introduce non-trivial biases when taking averages. This is indeed related to the effect known in sociology as the  \emph{friendship paradox}, first described by Scott L. Feld \cite{feld91friends}.

\emph{Why do your friends have more friends than you do?} Although it might seem surprising at first, this happens to be true for most of the people. Indeed, people with a lot of friends are much more likely to be friends with you than people with very few friends. In a sense, people with fewer friends (with lower connectivity) are harder to reach. As a consequence, although the average connectivity over a group of people might take a certain value, those people with higher connectivity than the average have a larger weight (they are considered more often) when taking averages over friends of friends. Of course, if someone has a number of friends larger than the average, the ``connectivity effect'' might not be large enough to produce the aforementioned situation. However, as shown in \cite{Ugander}, while the overall average number of friends in the studied social network is just $190$, one needs to have more than $800$ friends in order to have more friends than the average among one's friends. The connectivity effect is not negligible or small at all; on the contrary, it is quite a big deal.

Is something similar going on in the case of Voronoi graphs? Indeed, larger faces (with bigger connectivity) are more likely to be reached sooner by the expanding region under consideration. Is this effect playing an important role in our problem? If this tendency of larger faces to be sampled more often is significant, it could produce an artificial increase on the computed average, causing the curvature to take negative values---a quick inspection of equation (\ref{BCW2}) shows that averages larger than 6 correspond to negative curvatures. However, this case is quite different from the one presented in \cite{Ugander}. There, the average value of the quantity considered (number of friends) is quite high, and it has a much wider range of variability. Under these circumstances, the effect of the connectivity becomes larger. In our case, however, the expected average value of the variable is around $6$, and the range of variability goes from 3 to values rarely higher than $12$. In this case, the connectivity effects are limited to smaller deviations. We need to find the way to test if this phenomenon has actually anything to do with the case we are studying, and if it does, we need to find a way to avoid it, so we can finally compute a good estimate of the curvature. In the next section we will see that, indeed, this effect has to be taken into account and the biased statistics that it introduces lie at the heart of the deviations observed in the computed curvature.

\subsection{Alternative samplings}

In an attempt to solve the two questions raised at the end of the previous section, we make use of the fact that, in Voronoi graphs, all vertices are tri-valent. This implies that, unlike that of faces, the connectivity of all vertices is the same. This makes vertices the perfect candidates to avoid bringing into the problem connectivity effects that could alter the averages. We propose, thus, to use an initial vertex (instead of an initial face) and expand to increasing degree neighbouring vertices (following the graph structure). The selected region is now determined by the vertices and edges contained in a neighbourhood tree of degree $n$, starting from an initial vertex and including all vertices in the graph that can be reached from that vertex by $n$ hops or less. This method causes no vertices to be preferred over others. However, a certain freedom still remains. Since the ultimate quantity we need to determine in order to compute the curvature is the average of $p$ over a set of faces, we still have to establish a criterion to determine, given a certain neighbourhood tree, which faces are  selected by it. One has to be careful when establishing such a criterion, since there are different possibilities, each of them with potential consequences.

For instance, one of the possibilities would be to only consider those faces which are completely surrounded by the tree (\emph{full-cell}), i.e., all of whose vertices are part of the neighbourhood tree. On the other hand, one could choose to include all faces that are  ``touched'' by the tree, i.e., all those faces which contain at least \emph{one vertex} from the selected subgraph. There are also all kinds of intermediate choices, like for instance including those faces which have half or more of their vertices in the neighbourhood tree (\emph{half-cell}). One could even choose to consider some weighed average, in which faces that are only partially contained in the tree contribute to the average with a weight given by the fraction of the face contained in the tree.

Each of these choices can have different consequences according, again, to the different connectivity of faces. For instance, if one chooses to include all faces that contain at least one vertex from the tree, then again it is expectable that larger faces (containing more vertices) be more likely to be reached soon. However, if one picks the opposite choice, and includes faces in the region only when they have been completely surrounded by the tree, then the expected effect would be the opposite, namely, that smaller faces be favoured---since larger faces are more difficult to completely surround than smaller ones. Can we actually check that this is what happens with real computations? Is there a particular choice, somewhere in the middle, that balances out these two effects so we can get a good estimate of the curvature?

\begin{figure}[ht]
\begin{center}
\includegraphics[width=0.328\textwidth]{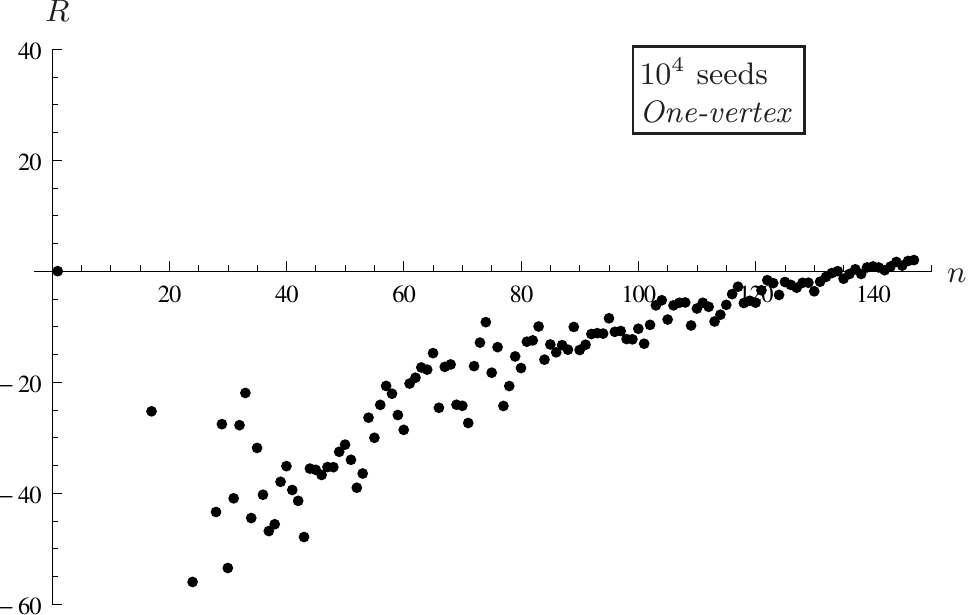}
\includegraphics[width=0.328\textwidth]{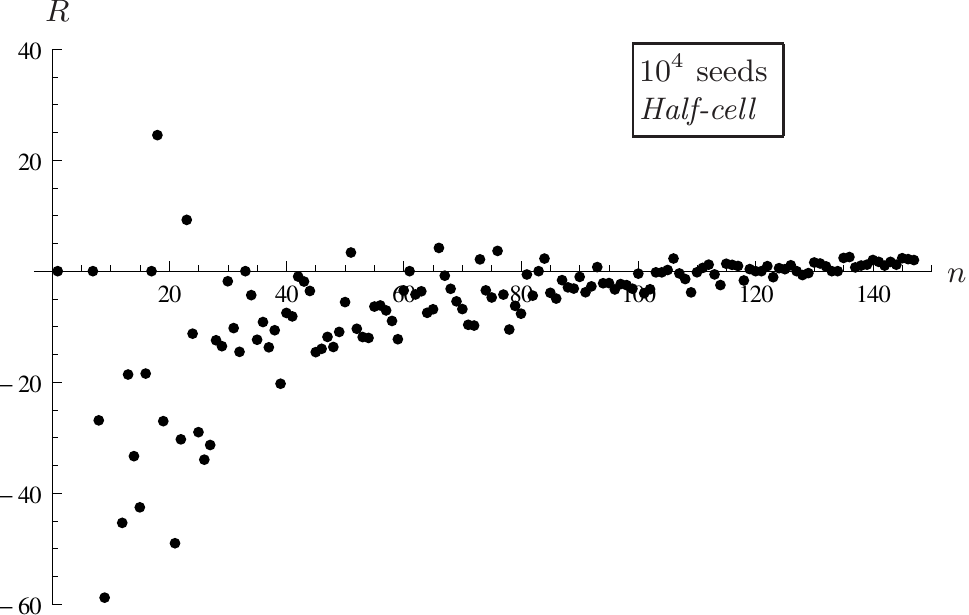}
\includegraphics[width=0.328\textwidth]{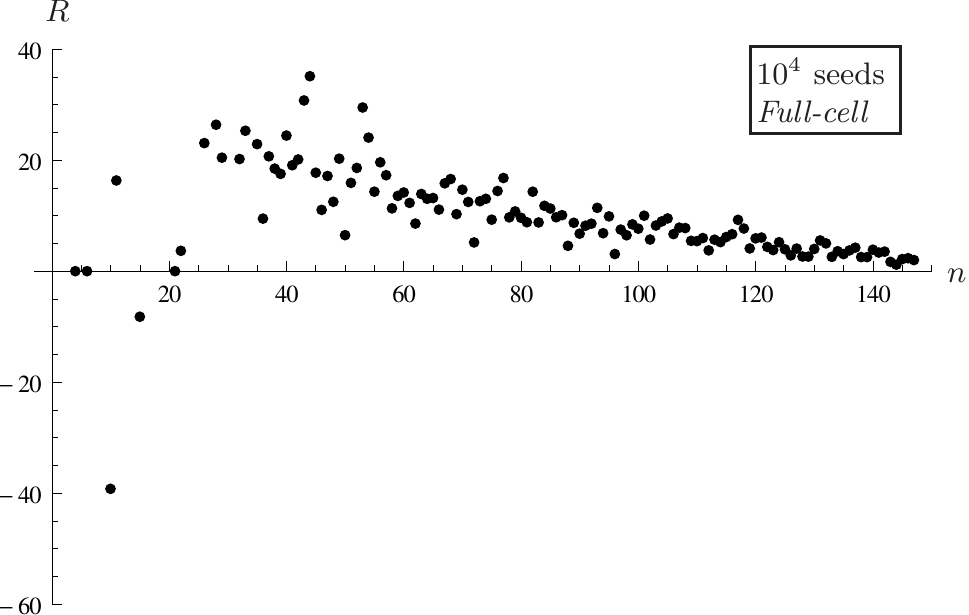}
\caption{This figure represents the values of the curvature of a unit sphere computed with the three proposed methods to select the faces considered in the sample region. On the left, the negative values of the curvature due to the expected effect of the bias favouring bigger faces is clearly observed. Analogously, on the right, we can see the opposite effect, as also expected. The plot on the center represents the \emph{half-cell} case where the curvature takes intermediate values.
\label{arbolfig}}
\end{center}
\end{figure}

To answer the first question, we simply have to compare the results for the curvature computed taking two different choices. We adapt our code to build regions based on neighbouring vertices (instead of neighbouring faces as we were doing before), and run our computations for the Voronoi diagrams again. The results obtained on a sphere for the three cases proposed above (completely surrounded faces, half-surrounded faces, and all faces reached by the tree) are shown in figure \ref{arbolfig}.

As we see, the differences are striking. For the \emph{one-vertex} case we still obtain negative curvatures, in agreement with the expected deviations. However, when considering only those faces which are completely surrounded by the tree, the effect gets inverted, also as expected, and we finally get positive curvatures. Although that is a step in the right direction, as we can see it does not make things much better, since curvature still varies monotonically (decreasing this time) and only reaches the correct value right at the end, when the whole sphere has been covered. This clearly shows that the effects of the different connectivity of faces are actually causing significant deviations on the computed averages, and are the cause for the large deviations observed in the computed curvature.

\begin{figure}[ht]
\begin{center}
\includegraphics[width=0.5\textwidth]{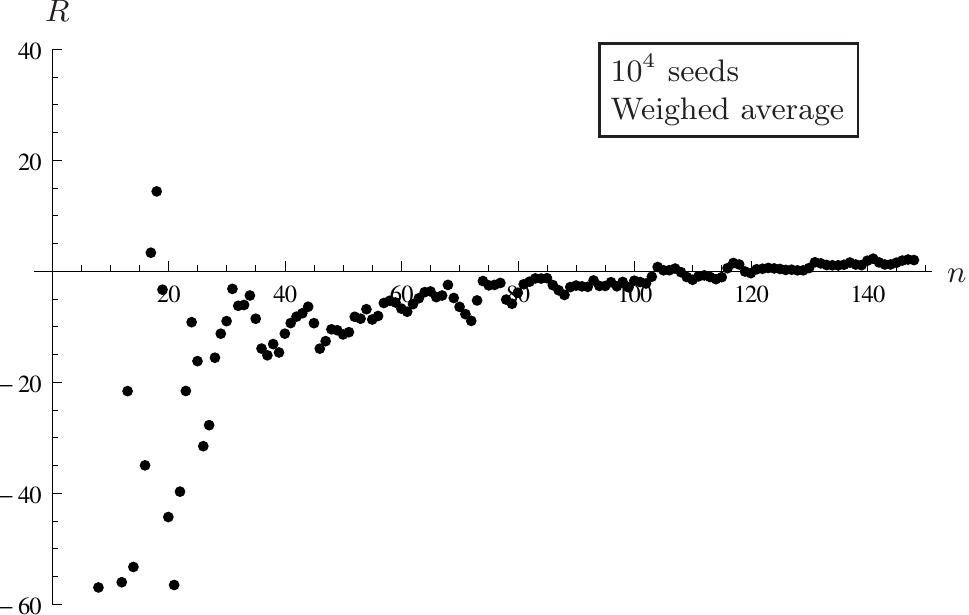}
\caption{The curvature computed with the method of weighed averages is represented in this figure as a function of the number of layers considered.
\label{prompesados}}
\end{center}
\end{figure}

We turn now to the second of the questions above. Can we avoid those effects and get a sensible enough estimate of curvature? In order to do so, we explore the intermediate choices. We saw already in figure \ref{arbolfig} the results for the \emph{half-cell} choice (selecting only faces with half or more of their vertices belonging to the tree). As we can see, curvature is still negative in this case for the most part, although there is a noticeable improvement with respect to the extreme \emph{one-vertex} case. We explore also the weighed average option, which in principle could seem like the most promising one. In this case, one is considering only the edges that are reached by the expanding tree, in a very precise sense. As shown in figure \ref{prompesados}, the curvature computed this way has a more stable behaviour, taking values significantly smaller---although still negative---than in the two previous cases. However, the improvements are not good enough to produce a positive---let alone accurate---value for the curvature of the sphere.

\subsection{Do finer graphs help improve statistics?}\label{statistics}

\begin{figure}[ht]
\begin{center}
\subfigure[]{
\label{random1K}\includegraphics[width=0.47\textwidth]{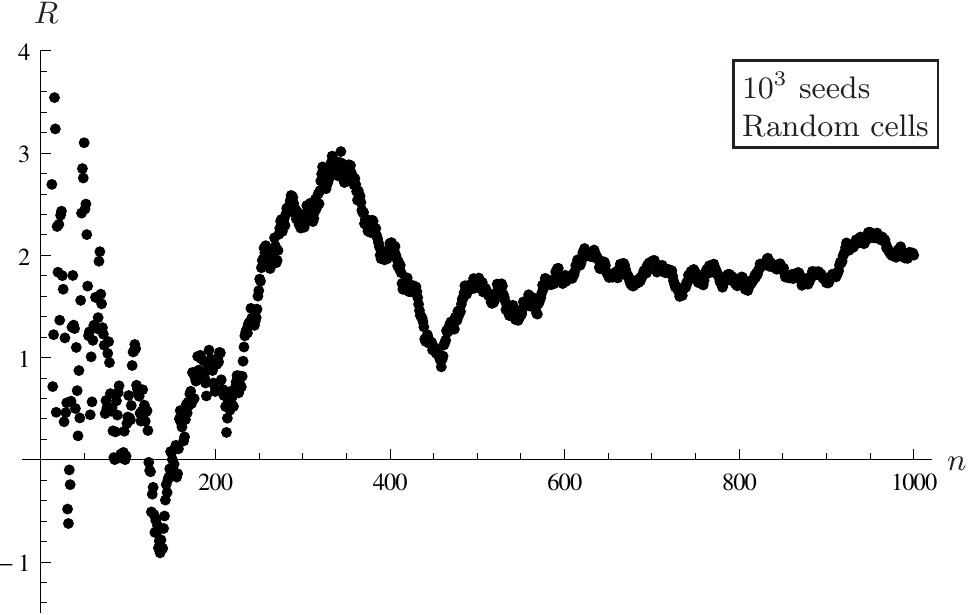}
}
\subfigure[]{\label{randomlarge}
\includegraphics[width=0.47\textwidth]{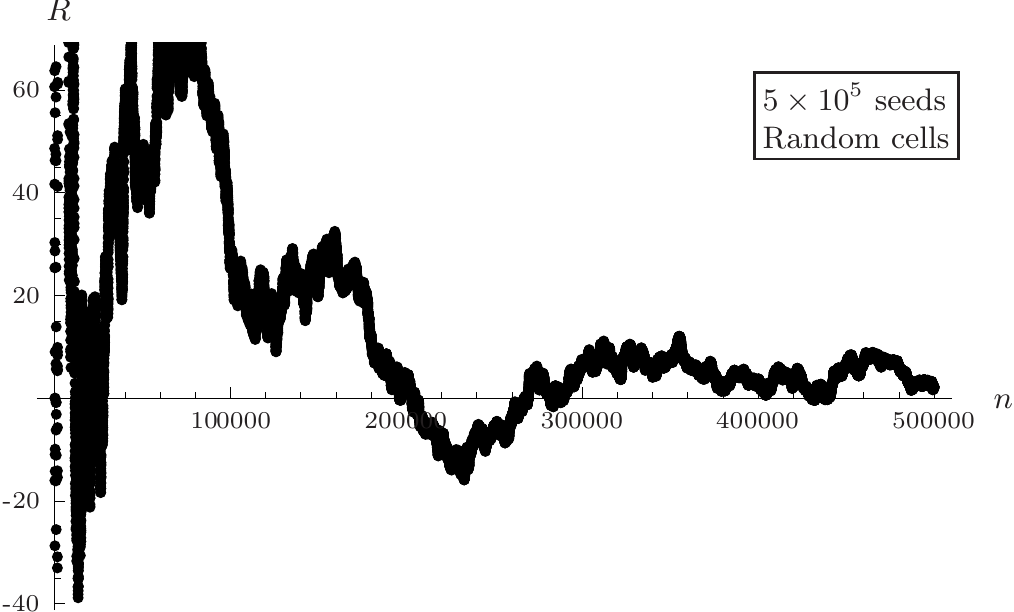}
}
\caption{The value of the curvature computed picking faces of the sphere at random is represented in this figure. We consider Voronoi diagrams with $10^3$ seeds in plot (a), and with $5\times 10^5$ seeds in plot (b).}
\end{center}
\end{figure}

We seem to be minimizing the effects of the different connectivities and, nevertheless, the computation of curvature is not giving good results. Is there something else going on? In order to make sure that there is no reminiscence of the bias on the choice of region, we take things one step further, to a completely unbiased choice: a random choice of faces. Since we are working on a sphere, all faces should be equivalent when representing the curvature of the surface, and we do not even need to worry about them being concentrated around a certain connected region. We simply pick faces from the sphere at random (using a  pseudo-random number generator), and keep computing the average as we keep adding new randomly picked faces to the subset of faces considered. Figure \ref{random1K} shows the results obtained when we do that for a sphere with a thousand seeds. As we can see, there actually is a point at which curvature seems to stabilize around the correct value of $2$. There are some fluctuations around that value, but on average the behaviour seems to be a good approximation to a constant. This stabilization happens only when about half of the faces in the sphere have been already picked, though. Can we get any better by adding more seeds and improving statistics?

We repeat the exercise for half a million seeds. In this case, the results are shown in figure \ref{randomlarge}. Much as it happened with the other plots, we observe that the behaviour seems to be ``scale invariant'', in the sense that adding more points does not mean that the curvature would stabilize earlier. Again, as a consequence of the topological nature of the computations, what matters is the fraction of the sphere considered, so only for about half the total amount of faces ($2.5\times 10^5$ in this case) the curvature stabilizes to a constant value. Furthermore, far from getting better, statistical fluctuations around the constant value of curvature (for the second half of the plot) are much larger in this case than in the previous one. By refining the graph, one gets worse, rather than better  statistics.

All this seems to point to the existence of some more fundamental issues. Let us go back to the equations in section \ref{section1}. We see, from equation (\ref{BCW2}) that curvature zero corresponds to an average number of $\bar p_0=6$ sides per face. Therefore, deviations of $\bar{p}$ from that number give us a measure of curvature. The quantity we are interested in, that is, the variable curvature is proportional to, is $\delta \bar p=\bar p-\bar p_0$. Furthermore, let us call $E_0$ the number of edges in a graph (with $F$ faces) such that this would have an average $\bar p_0$, and let us define $\delta E=E-E_0$. With that, and taking into account that each edge is shared by two faces, we can write:
\be
\label{delta1}
E=\frac{\bar p\ F}{2}=\frac{(\bar p_0+\delta \bar p)F}{2}=E_0+\frac{\delta \bar p\ F}{2}= E_0+\delta E\,\,\Rightarrow\,\, \delta E=\frac{\delta \bar p\ F}{2}.
\ee

Now, if we substitute the expression $\bar p=\bar p_0+\delta \bar p$ in equation (\ref{chi}), we find
\be
\label{delta2}
\chi=F\left(1-\frac{\bar p_0+\delta \bar p}{6}\right) = -\frac{\delta E}{3}\,\,\Rightarrow\,\, \delta E=-3\chi.
\ee
In other words: the \emph{deficit} of edges---the difference between the actual number of edges and the number required in order to get an average $\bar p_0=6$---is a constant value ($-6$ in the case of a sphere). It does not depend on the total number $F$ of faces; it does not vary when refining the graph by adding more seeds to the Voronoi diagram. This is one more consequence of the topological nature of the equations used in \cite{Bombelli:2004si}. But this one has very important consequences. Namely, the quantity we are trying to measure, $\delta \bar p$, is inversely proportional to the total number of faces (the number of seeds of the Voronoi diagram). From (\ref{delta1}) and (\ref{delta2}):
\be
\delta \bar p=\frac{2\ \delta E}{F}=-\frac{6\chi}{F}\propto\frac{1}{F}.
\ee

One could argue, at this point, that according to equation (\ref{BCW1}) this inverse proportionality with $F$ would cancel out with the factor $F$ appearing there. Indeed, we can rewrite the equation for the curvature as:
\be
\label{BCW3}
R=4\pi\frac{F}{A_S}\left(1-\frac{\bar p}{6}\right)=-4\pi\frac{F}{A_S}\frac{\delta \bar p}{6}\propto\frac{F}{A_S}\frac{1}{F}.
\ee
After all, since we are working on the unit sphere, the curvature is a constant, and does not depend on the total number of faces of the Voronoi diagram\footnote{Another way of looking at it would be to use the expression (\ref{BCW2}) instead, and work with constant density $\rho$ of faces, in which case, obviously, the area of the sphere increases with the number of faces, and therefore the curvature decreases accordingly.}---as actually observed in all our calculations when averaging over the whole sphere. However, the first $F$ factor is a global factor that multiplies both $\delta \bar p$ and its fluctuations. Therefore, when considering the quotient between the computed curvature and its fluctuations, the global factor $F$ drops out, and what matters is the value of $\delta \bar p\propto 1/F$.

We can ask, on the other hand, what happens to the fluctuations when we keep adding more seeds to improve the statistics. The standard deviation $\sigma_{\bar p}$ of $\bar p$ decreases as\footnote{Notice that $\sigma_{\bar p}$ refers to the standard deviation of the distribution of values of the average $\bar p$ taken over all subsets of $n$ faces of the total $F$ faces in the diagram ($n$ being a fixed fraction of $F$), as opposed to the standard deviation $\sigma(p)$ of the numbers of sides of the faces in one of those subsets.}:
\be
\sigma_{\bar p}\propto\frac{1}{\sqrt{F}}.
\ee
Therefore, the relative dispersion is actually proportional to the square root of the total number of faces:
\be
\frac{\sigma_{\bar p}}{\delta \bar p}\propto\frac{1/\sqrt{F}}{1/F}=\sqrt{F}.
\ee

By refining the graph---by adding more seeds to the Voroni diagram---we are actually making statistics worse! This is a direct consequence of the deficit $\delta E$ of edges being a constant, independent of $F$, which in turn is, as commented above, a consequence of the topological nature of equation (\ref{BCW2}).

Our main conclusion is, therefore, that while the proposal made by Bombelli \emph{et al.} in \cite{Bombelli:2004si} seems at first to be based on reasonable premises, when performing actual computations we found it to be fundamentally flawed. We find this situation  challenging, and the questions raised by this framework very interesting, so further research is needed in order to solve these problems, as well as to extend all these considerations to the more interesting---and more complicated---case of three dimensions.

\section{Conclusions and outlook}\label{section4}

The problem of reconstructing a continuous geometry starting from a discrete, more fundamental combinatorial structure, like a graph, is interesting for a wide range of research fields, from biology to physics through social sciences. In the case of LQG---theory whose Hilbert space is constructed using wave functions defined over graphs---the solution to this problem could provide interesting hints on the construction of semiclassical states, moving toward a connection with classical solutions of the Einstein equations.

In this article we have discussed and implemented the method proposed by Bombelli \emph{et al.} \cite{Bombelli:2004si,Bombelli:2009bb} to compute the curvature of a manifold from an abstract Voronoi graph associated to it. By making use of some topological arguments involving the Gauss-Bonnet theorem, a method to statistically compute the curvature in terms of the average number of sides of the faces in the graph is suggested. We tested this method for the simplest geometries: the unit sphere (constant positive curvature) and the plane (zero curvature). Unfortunately, we found highly unsatisfactory results for the value of the curvature in both cases.

A closer analysis showed that the theoretical foundations of the proposed method are consistent, even when taking into account the particular topology of the sampled subgraphs and the effect of considering regions with boundary. However, there are subtle issues with the proposed prescription to build the sampled regions, which favours the selection of faces with a higher number of sides, introducing a statistical bias in the computations---much in the way of the so-called \emph{friendship paradox} \cite{feld91friends}. The effects due to this bias turn out to completely dominate the computations of curvature, rendering the method inadequate for any practical purposes.

In an attempt to avoid the problem with biased statistics, we explored several ideas. We introduced an alternative way of carrying the graph sampling, based on neighbouring vertices---which present a uniform distribution in their degree of connectivity, for all of them are tri-valent in the type of Voronoi graphs considered---rather than on neighbouring faces. While the results were significantly modified---proving the relevant role that statistical biases play in this context---, we found it difficult to propose a completely unbiased method using this system, as certain arbitrary choices need still be made. The reduced degree of bias we were able to achieve remains the main factor dominating computations.

As a second alternative, we performed the computations by choosing the sampled faces at random. While this procedure seems to finally avoid the problems with the bias, a more fundamental issue was exposed by these computations. The topological nature of the considerations leading to the method proposed for the computation of curvature, turn out to have unwanted consequences. Namely, that considering finer graphs does not produce better statistical samples, for all that matters is the fraction of the sphere covered by the sampled region, and not how many faces that region contains. Even worse, rather than improving the statistics, the relative dispersion in the curvature gets larger the finer the graph, since it grows proportional to the square root of the total  number of faces of the Voronoi diagram. This is the main impediment for the studied method to produce sensible results.

Several approaches to the study of the transition from discrete to continuous structures remain to be explored. Motivated by the lessons learned from the analysis presented here, a more careful study of the statistics of the problem should be carried out. In particular, not only the information contained on averages and the biases that affect their values need to be considered, but also the much richer information contained in the different statistical distributions involved in the problem, in the dispersions and their behaviour with refining/coarse-graining of graphs, and in the relation with renormalization procedures (in line with well-known studies in spin-foam models, causal dynamical triangulations and Regge calculus). On the other hand, the use of Voronoi diagrams is well-suited for the study of LQG states in that their vertices (in 3 dimensions) are 4-valent and, therefore, among the simplest non-trivial structures containing information about volumes in that context. Then, one of the most important remaining challenges is the non-trivial extension of this framework to include 3-dimensional manifolds.  We leave the study of these approaches and extensions for future research.

\section*{Acknowledgements}

We thank Etera Livine and Simone Speziale for interesting comments and suggestions. We also thank Enrique F. Borja and David Jaramillo for several discussions during the initial stages of this project. This work was supported in part by grant NSF-PHY-1305000 and funds of the Hearne Institute for Theoretical Physics and by the Spanish MICINN research grants FIS2009-11893, FIS2012-34379, ESP2007-66542-C04-01, AYA2009-14027-C05-01 and AYA2011-29833-C05-01. IG is supported by the CNPq-Brasil.


\end{document}